**ORIGINAL ARTICLE**

# Evaluating the effectiveness of sonification in science education using Edukoi

Lucrezia Guiotto Nai Fovino[1] · Anita Zanella[2] · Luca Di Mascolo[3,4,5,6] · Michele Ginolfi[7] · Nicolò Carpita[1] · Francesco Trovato Manuncola[1] · Massimo Grassi[1]

**Abstract**
Science, Technology, Engineering, and Mathematics classes are mainly taught using visual supports. However, the advancement of technology and the increasing efforts to equip schools with digital instrumentation have opened up the possibility of exploring new teaching avenues, such as sonification. We explored the efficacy of sonification in education using a novel interactive tool, Edukoi, in the context of astronomy, which is predominantly disseminated through spectacular images, animations, and visuals. Edukoi is a motion-sensing sonification tool that converts images to sound in real-time for educational applications. Our study, conducted with nearly 150 middle-school students, included a preliminary questionnaire investigating the perception, engagement, and motivation of students towards science; two sessions dedicated to testing Edukoi and assessing the potentiality of the software for the recognition of the colour and the shape of real and sketchy images; and a final second administration of the questionnaire to capture a possible beneficial effect of the use of the tool in the engagement towards science. Results showed the effectiveness of Edukoi in colour recognition and reasonable efficacy in shape identification. Although the questionnaire did not reveal an increment in science engagement over the time of the study, oral feedback from the students was positive. Edukoi presents a possible alternative teaching aid, potentially benefiting diverse learners, including the visually impaired. Further developments of the software are needed to enhance its effectiveness in conveying more complex features such as composite colours or shapes.

## 1 Introduction

Science, Technology, Engineering, and Mathematics (STEM) classes are mostly taught using visual supports such as images, graphs, charts, and equations in digital and analogic form. Visual displays are valuable tools to represent datasets, reason about measurements, and process information. However, they also have some limitations. For example, they privilege students who prefer learning visually and penalise those who do not (e.g. more aurally oriented students, blind and visually impaired students [1]. This could influence the range and diversity of students pursuing scientific studies and careers.

In recent years, the advancement of technology and the increasing efforts to equip schools with digital instrumentation have opened up the possibility of exploring new teaching avenues. In particular, the growing circulation and capabilities of auditory displays create opportunities to complement visual displays, inviting educators to expand their teaching modalities, reimagine traditional image-centric learning tools, incorporate musical skills, and use

✉ Lucrezia Guiotto Nai Fovino
lucrezia.guiottonaifovino@studenti.unipd.it

1   Department of General Psychology, University of Padova, Padua, Italy
2   Istituto Nazionale Di Astrofisica, Vicolo Dell'Osservatorio 5, 35122 Padua, Italy
3   Astronomy Unit, Department of Physics, University of Trieste, Via Tiepolo 11, 34131 Trieste, Italy
4   INAF - Osservatorio Astronomico Di Trieste, Via Tiepolo 11, 34131 Trieste, Italy
5   IFPU - Institute for Fundamental Physics of the Universe, Via Beirut 2, 34014 Trieste, Italy
6   Laboratoire Lagrange, Université Côte d'Azur, Observatoire de La Côte d'Azur, CNRS, Blvd de L'Observatoire, CS 34229, 06304, Cedex 4 Nice, France
7   Dipartimento Di Fisica E Astronomia, Università Di Firenze, Via G. Sansone 1, 50019, Sesto Fiorentino, Florence, Italy

digital tools to support inclusive learning opportunities [2]. Introducing sonification in the school curricula and science classes is now possible.

Sonification is defined as the process of translating data into non-speech sound [3]. In the last decades, there has been a growing interest in using sonification to represent scientific data for research, outreach, artistic, and educational purposes [4]. Introducing sonification in the school curricula and adopting it as a teaching tool in the classroom, in tandem with more traditional visual methods, would make science teaching multimodal and multi-sensory. Using educational resources that offer multiple sensory and stylistic channels of accessing information and content, engaging the learners, and interacting is highly recommended in the framework of the Universal Design for Learning [5], namely the design and use of resources that are suitable for everyone, without needing ad hoc arrangements [6]. This could be achieved by introducing sonification in STEM classes, as it would allow blind and visually impaired students to access science education at the same level as sighted students [7], improving their engagement in the classroom [8] and also effectively creating a common ground for inclusion among people with diverse learning preferences (e.g. visually or aurally based) and abilities [9, 10]. For instance, it has been shown that students with dyslexia or autism might benefit from alternative learning modalities [5]. For many students, sound could promote a more well-rounded understanding of scientific phenomena than what would be possible through visual presentations alone [11]. Several studies highlight that allowing students to personalise their learning experience and, to some extent, take control of the learning framework is one path to educational success (e.g. [12–14]). For instance, these works show that presenting mathematical concepts in several representations at once can improve understanding. In addition, using sonification in schools might also help students who are not interested in scientific disciplines in the first place to get engaged and possibly overcome the so-called "science anxiety" that is perceived when more traditional methods are used in classes [15, 16]. Finally, although sonification is becoming an acknowledged method for data representation and communication, its uptake has been slow for a number of reasons. Ballora and colleagues [11] argue that this is partially due to the fact that relatively little work has focused on introducing auditory displays in education and learning. Making sonification an alternative tool at school would serve the long-term goal of enhancing the perception of scientific data, allowing students to consider science as a subject that can be understood not only through seeing but also by hearing [11].

Student attitudes towards science have been a focal area of research in science education for decades (for a recent review, see [17] and references therein). However, the impact of alternative teaching tools that involve the use of sonification on students' engagement in a regular school setting still remains largely unexplored. Evaluations of the effectiveness of multi-sensory representations of scientific data carried on with non-specialist audiences in informal educational settings such as museums or planetariums are encouraging [9, 10, 18–21]. Such studies have shown that the use of sound in informal STEM contexts is highly enjoyable and engaging for the audience and that multi-sensory presentations of knowledge produce a more educated and engaged community than the traditional "specialist" communication model [11]. However, most of the visitors of museums and planetariums (and other similar informal education environments) already have an interest in science that leads them to visit the exhibition or the show, making them a possibly biased audience. It is therefore important to assess whether sonification has a beneficial impact in formal education environments, such as schools, and whether it helps to engage more, and also more diverse students during science classes. For example, blind and visually impaired students are often hindered in their education by the massive use of visual representations of data, such as graphs, charts, and diagrams. Sonification could offer a viable alternative, potentially diminishing the barriers posed by non-accessible visual materials [22] and allowing BVI students to access information autonomously [23]. Introducing sonification in school curricula could equip BVI students with the confidence to use advanced sonification techniques later on for work purposes, including scientific research [24].

Our study leverages previous research conducted in informal educational environments and aims to test the effectiveness of sonification during science classes at school using Edukoi [25], an adaptation of Herakoi, a software for outreach purposes [26] (see paragraph 1.1). We aimed at achieving the following main goals: (1) assessing the effectiveness of sonification in conveying punctual information (colour and shapes) with a minimum amount of training; (2) investigating whether the use of sonification during science education at school changes the engagement and attitude of students towards such disciplines; (3) understanding whether the use of sonification during science classes changes the perception of students about accessibility of STEM education, namely about the possibility for blind individuals to pursue scientific studies and careers. This study only focuses on the first two questions, leaving the third point to a future publication.

To achieve the aforementioned goals, we carried out a study with about 150 middle-school students from an Italian public school.[1] Among the many STEM disciplines that we could choose, we decided to focus on astronomy and use images of galaxies in our experiment. Our choice of subject is motivated by three main reasons. Astronomy is typically regarded as a visual science, and it is mostly disseminated and taught through spectacular images, animations, and graphs. However, this can be misleading for the non-experts that interact with the discipline, since most of the matter in the Universe does not produce or absorb light, and even that which does is usually not visible to the human eye (e.g. too faint or not in the visible frequency range, e.g. [10]). For instance, galaxies are too faint and distant to be seen with our eyes; most of the light that they emit is not in the frequency range that the human eye can capture; and finally, most of the matter that galaxies are composed of seems not to emit light at all (i.e. the so-called "dark matter") and hence it cannot be directly seen. Sophisticated telescopes and computers are required to turn such astronomical data into visible images. There has been an emerging research interest in converting astronomical phenomena into sound, instead of images, that has risen in the last decade (e.g. [19, 24, 27]). Some of the key features that are striking when looking at galaxy images are their colours and shapes. Colours in astronomy are traditionally related to some physical properties of galaxies (e.g. their age, metal and dust content) that allow astronomers to investigate galaxy formation and evolution, but they are not necessarily linked to the "real" colours that we normally perceive. This allowed us to discuss with the students misconceptions related to visual and sound representations of data. In addition, the software we used for our experiment was created to explore astronomical images, although it can be used for other purposes.

---

[1] Anecdotally, during the software development, we were contacted by some science teachers that were searching for ways to teach their blind and visually impaired students.

In our study, we tested whether the colours and shapes of galaxies could be distinguished through sound only, by using the interactive sonification tool Edukoi. We asked students to explore some geometric shapes with basic colours first, and some sketchy and realistic astronomical images later. With this work, we also aim to let the students understand that teaching and approaching astronomy—and science at large—through visual displays is just one of several possible choices. In many cases, as for galaxy images, visualisations are simply one of the possible representations of numerical data.

### 1.1 Our sonification tool: Edukoi

Edukoi is an adaptation of the open-source software Herakoi [26]. They both are motion-sensing sonification tools that convert images into sound in real-time by using the publicly available MediaPipe Hand Landmarker model [28] to detect hands' landmarks in an image, tracking the keypoint localisation of 21 hand-knuckle coordinates. According to the MediaPipe documentation, their model was trained on approximately 30,000 real-world images and on several synthetic hand models imposed over various backgrounds. The model bundle combines the hand landmarks detection model with a palm detection model, resulting in a nested detection algorithm with increasingly refined modelling of the hand coordinates. In fact, the palm-tailored algorithm locates hands within any input image, thus defining a region of interest within which the hand landmarks detection model identifies specific hand landmarks. The MediaPipe Hand Landmarker task operates on image data that can be either static data or a continuous stream. We use it to track the position of the hand(s) of the user(s) in real-time in the scene observed by a webcam connected to the user's computer [28]. The model records the coordinates of the user's hands, which are then re-projected onto the pixel coordinates of the chosen image. The system offers two options for defining the "touched" area, based on the user's needs. The first option is to define a square region, with the side length determined by a customisable number of pixels. This square is centred around the coordinates of the user's index finger. The second option is to define a rectangular region, with its shape determined by the coordinates of both the index finger and the thumb. With the square region option, the user interacts with a fixed area around their index finger. They can adjust the size of this area by customising the number of pixels that make up the side of the square. This provides a precise and controlled interaction suitable for certain applications. On the other hand, the rectangular region option allows the user to interact with the image in a more dynamic and adaptive way. By using the coordinates of both the index finger and the thumb, the system creates a flexible rectangle. Depending on how far apart the index finger and thumb are, the size and aspect ratio of the rectangle change accordingly. This enables the user to explore different scales of the image, zooming in and out naturally as they interact.

The visual properties of the "touched" pixel are then converted into the sound properties of the selected instrument, which can be chosen from a virtual MIDI keyboard [26]. The process of converting visual properties into sound properties is highly customisable, allowing the user to modulate various instrument characteristics according to their preferences, enabling a wide range of sound expressions. For instance, as in the current default mode of the software, the user can map the pitch of the sound to the chromatic hue of the "touched" area and adjust the intensity of the sound based on its saturation.

The current default image-to-sound mapping used in Herakoi is consistent with the most popular ones used for the sonification of astronomical data [24, 27]. It employs a physical translation of the concept of wavelength and frequency for connecting any specific colour in the target image to the pitch of the output sound (i.e. colour–pitch, with red being low-pitched and blue being high-pitched). The colour scale is based on the hue parameter from an HSB model of the input image, truncated at 80% of the total hue scale to avoid potential issues with the cyclic nature inherent to the HSB representation. Given the specific choice for Herakoi to generate the sound information in the form of MIDI messages, the pitch values are discretised over the chosen tuning range. The brightness parameter is instead mapped to the amplitude of the output sound (i.e. dimmer and brighter pixels correspond to quieter and louder sounds, respectively) by means of a linear mapping from the full brightness scale to the selected dynamic range for the sound amplitude. Such an approach is best suited to engage the public and offer an entertaining experience, but it is not optimal to extract and remember information on colours and hues from the images that the user explores. In fact, associating colours with pitches requires long-term memory of both pitches and the pitch-colour association. The ability to identify the pitch chroma of a note presented in isolation is called "absolute pitch". Absolute pitch is extremely rare in the population (estimated less than 1 in 10,000 people, Deutsch 2013). People without absolute pitch and/or without extensive musical training find it difficult to compare and remember sounds of different frequencies. Memorising information and remembering it without the continuous need of training sessions is key in educational environments. For these reasons, we used a different colour-to-sound mapping in Edukoi (see Sectio 2 for details).

The paper is organised as follows: in Section 2, we describe the sonification tool, mapping, testing users, and the experimental procedure. In Section 3, we discuss the implementation and assessment of our experiment. In Section 4, we report the results, and in Section 5, we discuss their implications. Finally, in Section 6, we summarise and conclude. This paper extends the conference proceedings presented at ICAD2023 [29].

## 2 Methods

The aim of our study was to let middle-school students (and teachers) discover sonification and understand whether it makes the learning of scientific disciplines more engaging. We also aimed to evaluate the effectiveness of sonification in conveying information about geometric shapes, colours, and the content of astronomical images while letting students understand that sonification makes astronomy (and science) accessible to blind and visually impaired pupils, possibly changing their current perspective. In the following, we describe the sonification tool, the methods used in our study, and the project timeline.

### 2.1 Sonification tool and adopted mapping

One of the most common sonification methods is called parameter mapping [30]. In parameter mapping, the sound characteristics (e.g. pitch, volume, or timbre) are tied to the properties of the data. Mapping sound to physical parameters involves a series of arbitrary choices. However, it has been shown that some mappings might be more effective than others in conveying certain types of information [27].

We chose to map the three red–green–blue (RGB) colours in an associative way. In order to enhance memorisation and ensure the correct identification of colours even after extended periods of non-use, mitigating the risk of needing a familiarisation process several times, we employed natural sounds to exploit the association between sounds and colours in our mapping. Specifically, the sound of a crackling fire was associated with the colour red, the sound of water bubbles with blue, and the sounds of birds and rustling leaves with green. This approach should facilitate the recollection of the sound-to-colour coding through the establishment of timbre-word associations. This association is also expected to be independent of the users' culture and level of visual impairment [17, 31]. One possible drawback of our mapping is that the RGB coding makes it difficult to analyse more complex colours (e.g. composite colours: magenta, yellow, etc.), as the theory of colour composition is not necessarily mastered by young students and it is not immediate to all. Switching to cyan–magenta–yellow–black (CMYK) colour mapping in future studies might make the identification of composite colours more intuitive at least for sighted students who had a chance to experience the subtractive colour theory (e.g. with tempera).

We also chose to spatialise the sounds to make colour identification easier when, for example, two colours are played together.[2] When listening with headphones or other stereo devices, the red is played by the left earpad, the blue by the right one, and the green by both pads.

---

[2] In general, when senses deliver contrasting spatial information, audition assumes the location delivered by the concurrent sense as correct. A typical case is the ventriloquist effect, in which the spatial location of a sound source is attributed to the spatial location of a visual stimulus [32]: although the sound of the TV comes from the loudspeakers, we perceive the voice of the person on the screen as originating from his/her mouth and not from the loudspeaker. A similar type of ventriloquist effect occurs when the sound source location contrasts with proprioceptive information about the position of our body parts [33]. In practice and in brief, it is unlikely that the spatial location of the sound affected the perceived position of the object the participant was exploring.

**Table 1** Summary of the sessions of our study

| Test phase | Time | Purpose | Evaluation tool | Number of students |
| --- | --- | --- | --- | --- |
| Q1 | Beginning of Feb. 2023 | Evaluate the perception of science | mATSI questionnaire | 138 |
| T1 | 17–18 Feb. 2023 | Colour and geometric shape recognition | Behavioural test | Colour task: 146<br>Shape task: 109 |
| T2 | 19– 20 May 2023 | Galaxy colour and shape recognition | Behavioural test | Colour task (sketchy images): 148<br>Colour task (natural images): 126<br>Shape task (natural images): 91 |
| Q2 | End of May 2023 | Evaluate the perception of science | mATSI questionnaire | 91 (of which 67 took also Q1) |

## 2.2 Testing users

We partnered with the middle school "Scuola Media Don Milani", located in Castiglione delle Stiviere (Mantova, Italy). Classes were voluntarily recruited with reference to science teachers. Seven classes adhered to the study for a total of about 150 students (two classes 1st grade, two classes 2nd grade, three classes 3rd grade). Students were between 11 and 14 years old.

In the various phases of the experiment, students participating in the study could be less than 150 because of sick leaves, individual technical difficulties in the testing session, or testing sessions occurring in particularly busy school periods (e.g. some classes could not take one part of the testing altogether because testing occurred simultaneously with the preparation of the end of the year exams). Among the students, one was completely blind and three had different disabilities (deafness, cognitive impairment, non-verbal autism). These students were assisted by a support teacher. Ten students had some mild learning impairments. Finally, about 10% of the students were not native Italian (e.g. born in China, Ukraine, India, and African countries), but most had good Italian proficiency. Everyone participated in the study to their best capability. Data were collected anonymously: each student was identified by a unique alphanumeric identification code.

## 2.3 Evaluation tools and work plan

The immediate objectives of our study were two: (1) testing the efficacy of the sonification tool for colour and shape recognition; (2) evaluating whether the use of Edukoi and sonification in science classes could have an impact on the students' enjoyment of scientific subjects. We assessed the colour and shape recognition with a behavioural experiment. We assessed the perceptions, engagement, and motivations towards science of the participants with the modified Attitudes Towards Science Inventory (mATSI) questionnaires [17, 31].

The study unfolded along four sessions, hereafter referred to as Q1, T1, T2, and Q2 (Table 1). In Q1 (beginning of February 2023), we administered the mATSI questionnaire to assess attitudes towards scientific subjects prior to our intervention in the classes. The questionnaire was administered by the teachers who gave little to no explanation about the study. The students were informed that they would receive information about the experiment later (during T1). A few days before T1, we also collected the semi-structured interviews with the school psychologist of the school. In T1 (February 17th–18th, 2023), we explained the study and introduced the concept of sonification. Successively, students run a set of behavioural tests using Edukoi to recognise colours and shapes through the exploration of basic geometric shapes depicted in various colours. In T2 (May 19th–20th, 2023), students run a second set of behavioural tests using Edukoi. These tests focused on colour recognition and extended also to the recognition of basic geometric shapes. T1 and T2 allowed us to investigate whether properties like colours and shapes can be recognised by using Edukoi. By repeating the experiment twice, with an interval of about 2 months between T1 and T2, we also assessed whether the information provided by the sonification (e.g. colour mapping, shape investigation strategies) could be remembered over a relatively long time. Eventually, in Q2 (end of May 2023), we conducted the second round of semi-structured interviews and administered the mATSI questionnaire a second time to re-assess the attitude towards scientific subjects and to capture attitude changes in comparison to Q1. Finally, in T1 and T2, we did not include a vision condition (i.e. a comparison condition) because not informative: in sighted students, the performance in a visual condition would have been at ceiling (see Fig. 1), whereas in BVI students, it would have been impossible.

## 3 Implementation and assessment

In the following section, we describe in detail the different testing phases (Q1, T1, T2, and Q2). The questionnaire and behavioural tests described below were conducted in the school's computer room with 20–25 students at the time. We also collected qualitative data regarding the perception of our tool and the best ways to implement it further. We conducted pre- and post-semi-structured interviews in collaboration with the school psychologist. Some responses are reported in Appendix 3, but full data collected in these interviews is not discussed here.

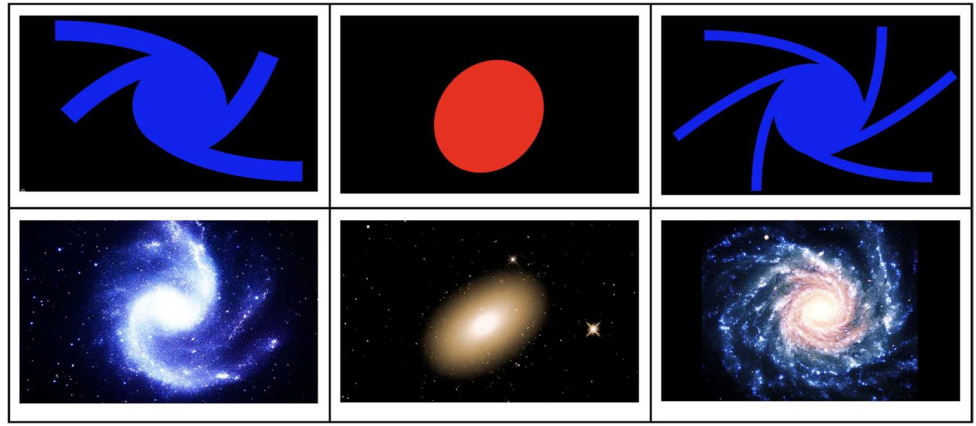

**Fig. 1** Sample of the sketches presented during the first colour-recognition task (first row) and of the real galaxy images used for the second and third tasks (second row) of the T2 session. In the first task, students had to recognise the colour of the shape, while in the second task, they had to recognise all colours present in each galaxy image, and in the third task the galaxy shape

### 3.1 Q1: First administration of the mATSI questionnaire and semi-structured interviews

Q1 was conducted during the first week of February 2023 by the teachers of the various classes participating in the study, and 138 students responded to the questionnaire. Teachers administered the mATSI questionnaire via a Google Form that each student could access during school hours. mATSI consists of 25 questions, and responses are given on a Likert scale from 1 (not at all) to 5 (very much). Specifically tailored for elementary and middle-school students, the mATSI captures critical information about student attitudes towards science. The mATSI measures several independent elements within the dimension of attitudes towards science, such as the "science anxiety" or "value/enjoyment of science". The mATSI questionnaire is a powerful tool for understanding the root causes of "science anxiety" and helping devise practical plans to assist students and foster a positive learning environment [34]. For our study, we translated the mATSI questionnaire that was originally written in English into Italian [17, 31].

The questions are grouped according to specific themes:

- Perception of the teacher (e.g. "Science teachers make science interesting"; "It scares me to have to take a science class")
- Anxiety towards science (e.g. "It makes me nervous to even think about doing science")
- Value of science to society (e.g. "Science is useful for solving the problems of everyday life"; "Most people should study some science")
- Self-confidence in science (e.g. "Science is easy for me"; "I do not do very well in science")

- Desire to do science (e.g. "Science is something which I enjoy very much"; "Sometimes I read ahead in our science book")

In Appendix 1, we report all the questions that we asked in the framework of the mATSI questionnaire.

### 3.2 T1: first empirical test of EDUKOI

T1 was conducted on February 17th–18th, 2023, by two of the authors (LG and AZ). A brief astronomy lesson about the work of the astronomers and the meaning of colours in astronomy, for example, to explore the difference between young (blue) and old (red) stars, preceded the sonification test. One hundred forty-six students took the test, but only 109 students participated in all its parts. Teachers were present but did not interact with the students and simply helped to keep the environment collaborative. We presented the study and introduced the sonification concept, briefly explaining to students how Edukoi works and the basics of the sonification mapping adopted by the software. We made clear to students that sound does not propagate in empty space; hence, the sounds they hear were arbitrary representations of data (in this specific case, colours). At this stage, students were allowed to familiarise themselves with the tool. Successively, the experimenter mentioned the structure of the experiment, namely that students would first use Edukoi to recognise colours and then to recognise both colours and shapes. In both cases, students were asked to explore basic geometric shapes in three possible colours.

The behavioural test was conducted using Edukoi version 2.0 (Sect. 2.1), installed on the laptop PCs of the school running Windows 10. Each student used a computer provided by the school and their personal headphones or headphones provided by the school. The screens of the laptops were covered with thick, opaque, nontransparent paper. The first task (colour recognition) was made by six trials in which three geometric shapes were presented in three possible colours.

The three geometric shapes were a circle, a square, and an equilateral triangle of identical size (in pixels per $cm^2$). We chose these shapes because they are part of the Italian school curriculum of the participants and we wanted them to familiarise with the tool before presenting more complex and unfamiliar shapes (see T2). Shapes were painted in green, red, or blue (RGB pure colours). During this task, students were asked to explore the shape with their hands and only recognise the colour of the shape. Trials were run in identical order by all students, and the order was randomised beforehand. At the end of the exploration, students wrote the response on a sheet of paper. All 146 students completed this task.

After the colour recognition task, students were asked to do a second task during which they had to explore an image of the sky and search for stars. The sky was black, and thus silent, while the stars were red or blue. This task focused on object search and colour recognition only. Due to a technical issue, the software was not loading the pictures correctly, and responses had to be disregarded. After the second task, students underwent a third task in which they were asked to recognise not only the colour but also the shape of the picture they were presented with. This task was made of eighteen trials in which, again, three geometric shapes were presented in three possible colours. Shapes and colours were the same as during the first task. Trials were run in identical order by all students, and the order was random. At the end of the exploration, students wrote the response on a sheet of paper. Because this task was run at the end of the testing session, not all students participating were able to complete all the trials of this task. In the analysis, we only consider the responses from the 109 students who completed this task. Overall, the session lasted 50 min, but the actual duration of the testing was approximately 30–35 min.

### 3.3 T2: second empirical test of EDUKOI

We ran T2 on May 19th–20th, 2023, in the computer room of the school (see previous section for a description of the apparatus). A brief astronomy lesson on galaxies and how to estimate their age (the difference between old and young galaxies, composed of old and young stars as explained in T1) introduced the test. In order to assess the students' learning from T1, in T2, no explanation on how to use Edukoi and on the adopted mapping was given prior to the test. Authors (LGNF, AZ, and MG) were conducting the study, and teachers helped keep the environment collaborative. In T2, we presented to the students two sets of images. The first set of pictures explored by the students consisted of twelve stylised sketches of galaxies, which could be elliptical or spiral, with varying quantities of arms and inclinations. The images were displayed in red or blue, with six red images and six blue images. The same image could be presented in blue and in red. In this task, green was not used because galaxies are often classified as "blue" (star-forming) or "red" (passive), as they are dominated by those two colours. In Fig. 1 (first row), we show a sample of sketches. The second set of images consisted of twelve images of real elliptical or spiral galaxies. As these were "natural" pictures, they had complex colour patterns and could include composite colours (not only pure red or pure blue). Edukoi plays composite colours following the additive colour synthesis (Sect. 2.2). For example, the purple colour, which is made of blue and red, sounds simultaneously as water bubbles (associated with blue) and crackling fire (associated with red). In composite colours, the intensity of each sound corresponds to the proportion of each colour, thus a reddish purple with a larger component of red than blue will sound with louder fire than water bubbles. We used images taken from the archives of the European Southern Observatory[3] and of NASA.[4] In Fig. 1 (second row), we show a sample of these images. The complete list is shown in Appendix 2.

In T2, students were asked to respond to three tasks. Tasks were preceded by a brief explanation of astronomical concepts, focusing on the differences between spiral and elliptical galaxies. We also mentioned the structure of the session, namely that we would first use Edukoi on some sketches and then on actual astronomical data showing real galaxies. As in T1, the screens were covered with thick, opaque, nontransparent paper.

In the first task, students were asked to explore the stylised sketches of galaxies (elliptical or spiral), and they were asked to write down the colour (red or blue) of the sketched galaxy. The images were presented in random order, totalling twelve trials. The order was identical for all students. One hundred forty-eight students provided valuable responses for this task. In the second task, the students were asked to explore the real images of spiral and elliptical galaxies. The twelve images were presented in random order, identical for all students. Students had to explore the image to locate the galaxy and identify all colours included in the picture. They were asked to check boxes for each colour they could hear (red, blue, and/or green). Since training them on the theory of colour composition was beyond the scope of the current study, we asked them to check boxes for each individual colour they could hear (red, blue, and/or green). We did not ask them what composite colour resulted from this superposition. When this task was completed, students were asked to explore the real images of the galaxies again, but this time in reversed order (from the last trial to the first one) and to write down the galaxy's shape (spiral or elliptical). Respectively, 91 and 126 students provided valuable responses for the colour and shape recognition task with natural images. Overall, the session lasted 50 min, but the actual duration of the testing was approximately 30–35 min. Not all students participating in T2 were able to complete all the tasks.

---

[3] https://www.hq.eso.org/public/images/

[4] https://www.nasa.gov/multimedia/imagegallery/index.html

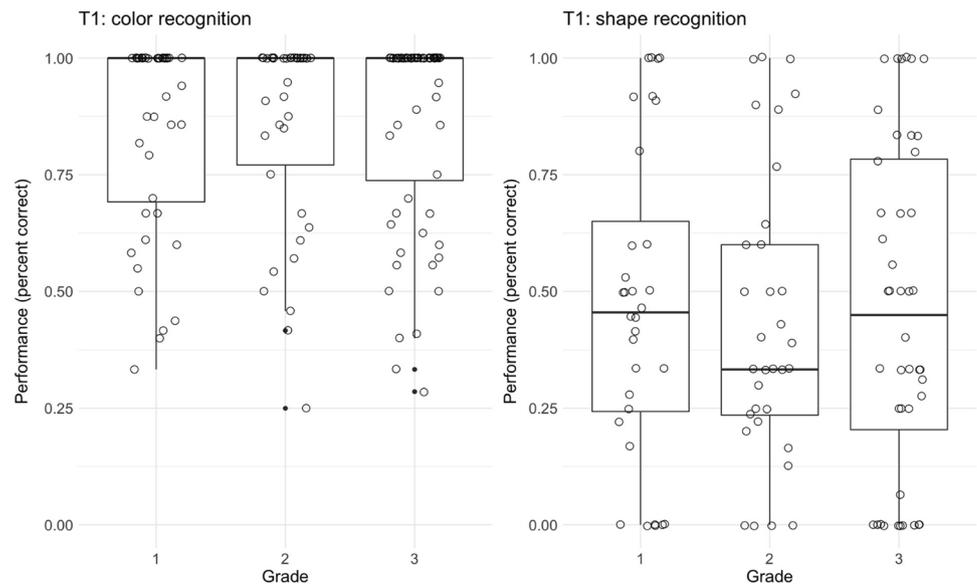

**Fig. 2** Performance of students in the testing phase T1 for colour identification task (left) and shape identification task (right). Results are represented as a function of the grade of the participants. Boxplots represent the performance of the three grades or equivalently of the three age groups. In each box, the black thick horizontal line marks the median, the edges of the box are the 25th and 75th percentiles, the whiskers are the interquartile range augmented by 50%, and the black symbols are outliers. Over each plot, we also represented the individual performances of the students

### 3.4 Q2: second administration of the mATSI questionnaire and semi-structured interviews

Q2 was conducted during the last week of May 2023, about 1 week after T2 and 91 students responded to the questionnaire. The semi-structured interviews were conducted in the days after T2, and the mATSI questionnaire was made accessible once again through a Google form. Only one class filled in the questionnaire during school hours and under the surveillance of a teacher, while the others filled it in at home. The choice of letting the students fill in the questionnaire at home was driven by the proximity of the end-of-term exams that took most of the time during school hours.

## 4 Results

Our experiment aimed to assess the effectiveness of sonification in tasks such as colour and shape recognition. We also aimed to verify whether the chosen mapping was effective and allowed students not only to understand information but also to remember it after some time (i.e. in our case 2 months) of inactivity. Finally, we investigated whether the use of sonification in the classroom could modify the perception and attitude of students towards science education.

### 4.1 Colour and shape recognition

The shape and colour responses of the students were coded as correct/incorrect and were analysed with a one-sample Student's *t*-test to compare the participants' colour and shape correct responses with those expected by chance. We also calculated Cohen's *d* (or standardised mean difference) to assess the magnitude of the effect size and the Bayes Factor (BF) to test the alternative hypothesis that the recognition performance was superior to the chance level. Finally, we calculated an analysis of variance (ANOVA) to investigate whether there were substantial differences in colour and shape recognition performances of students in different grades, corresponding to different age groups. Cognitive development from 11 to 14 years is fast and adolescence involves significant changes in several dimensions of cognition [35]. The ANOVA explored whether these changes had a significant contribution to the performance of the students.[5]

---

[5] Note that although our data were not normally distributed with homogeneous variance, both *t*-test and ANOVA are robust even when distributions of the data violate the distribution assumed by the test [36–38]. We also calculated the statistical analysis with non-parametric tests (e.g. Wilcoxon Signed-Rank Test instead of *t*-test or Kruskal-Wallis test instead of ANOVA) and found the same results. In the current version of the paper, we kept the parametric tests because we think they are more familiar to readers. The non-parametric alternatives were added to the analysis scripts that are available in the data/script repository of the study.

Table 2  Summary of the results about colour and shape recognition performed during T1 and T2

|  | Colour recognition | | | Shape recognition | |
| --- | --- | --- | --- | --- | --- |
|  | T1 (basic shapes) | T2 (sketchy galaxies) | T2 (real galaxies) | T1 (basic shapes) | T2 (real galaxies) |
| Number of participants | 146 | 148 | 91 | 109 | 126 |
| Average performance per student | 86% | 93% | 78% | 45% | 64% |
| Median performance per student | 100% | 100% | 80% | 41% | 64% |
| Interquartile range | 71.25–100% | 25–75% | 69.4–91.6% | 23.52–66.6% | 50–75% |
| $t$-test $t$ | 30.70 | 35.48 | 8.15 | 3.83 | 9.13 |
| $t$-test $p$ | <0.001 | <0.001 | <0.001 | 0.0002 | <0.001 |
| Cohen test $d$ | 3.59 | 4.12 | 1.20 | 0.52 | 1.15 |
| Bayes factor | 5.180432e+61 | 1.920099e+70 | 4177246846 | 90.3 | 4.177327e+12 |
| ANOVA $F$ | 0.11 | 2.89 | 18.79 | 0.07 | 3.88 |
| ANOVA $p$ | 0.88 | 0.0584 | <0.001 | 0.92 | 0.0232 |

### 4.1.1 Testing phase T1

One hundred and forty-six students provided responses for the two colour-recognition tasks of the geometric shapes. The performance (i.e. average proportion of correct responses) of students in identifying colours during the testing phase T1 is shown in Fig. 2 (left). We represented separately the results of students in different grades (11–12 years old in 1st grade, 12–13 years old in 2nd grade, and 13–14 years old in 3rd grade). The graph shows a good performance of the students—almost at the ceiling—with an average colour identification performance of 86% and a median value of 100%. Overall, the performance of the students was much higher than chance (i.e. 33%): $t(145) = 30.70$, $p < 0.001$, $d = 3.59$, BF = 5.180432e+61. There was no difference in performance across the three grades of students participating in the experiment; this was revealed by the analysis of variance that compared the performances of the three grades: $F(2, 143) = 0.11$, $p = 0.88$ (Table 2).

The second task of T1 included the recognition of geometric shapes, in addition to the recognition of their colour. Due to lack of time, not all the students completed this task. For our analysis, we only consider the one hundred and nine complete tests. Figure 2 (right) shows the performance (i.e. success rate) of students in identifying geometric shapes during the testing phase T1. It shows a performance above chance level (33%), with an average rate of correct responses for each student of 45%.

We computed a one-sample $t$-test against chance (i.e. 33%) and it was significant: $t(108) = 3.83$, $p < 0.001$, $d = 0.52$, BF = 90.39. We also tested whether there was a different performance across the three grades with an analysis of variance. As for the colour recognition task, the analysis revealed that the performance of students did not depend on the grade they were in: $F(2, 106) = 0.07$, $p = 0.92$. Overall, results show that participants had a performance higher (or much higher) than a chance for shape and colour recognition. We observed no difference in the performances across grades.

### 4.1.2 Testing phase T2

One hundred and forty-eight students responded to the colour recognition task of the stylised sketches of galaxies. The students' responses were highly above chance: $t(147) = 35.48$, $p < 0.0001$, $d = 4.12$, BF = 1.920099e+70. There was no significant difference in performance between the grades in this task: $F(2, 145) = 2.89$, $p = 0.0584$. The results are reported in Fig. 3.

We also analysed the responses given for the real images of the galaxies. For this task, we collected a total of 91 full datasets (i.e. students that provide both the responses for the colour and the shape recognition with the real images of galaxies), plus another 35 datasets about colour recognition only. Because real images of galaxies included many colours and hues and not simplified RGB colours only, students could in principle recognise all the RGB components: the red, the blue, and the green. However, in order to do so, students had to explore the image in all its parts. For each student, we assessed whether s/he recognised the red, blue, and green colours, then we averaged the three responses into a single colour performance index. In practice, for a given image, the student recognising the presence of all colours scored 1, the student recognising two colours out of three scored 0.66, that recognising one colour only out of three scored 0.33 and that recognising no colour scored 0. In practice, this colour index ranges from 0 (no colour has been recognised) to 1 (all colours have been recognised). The index was calculated separately for each student and image and successively averaged for each student. We calculated a one-sample $t$-test to see whether the average performance of students was above 0.66 (i.e. the recognition of more than two colours). Students

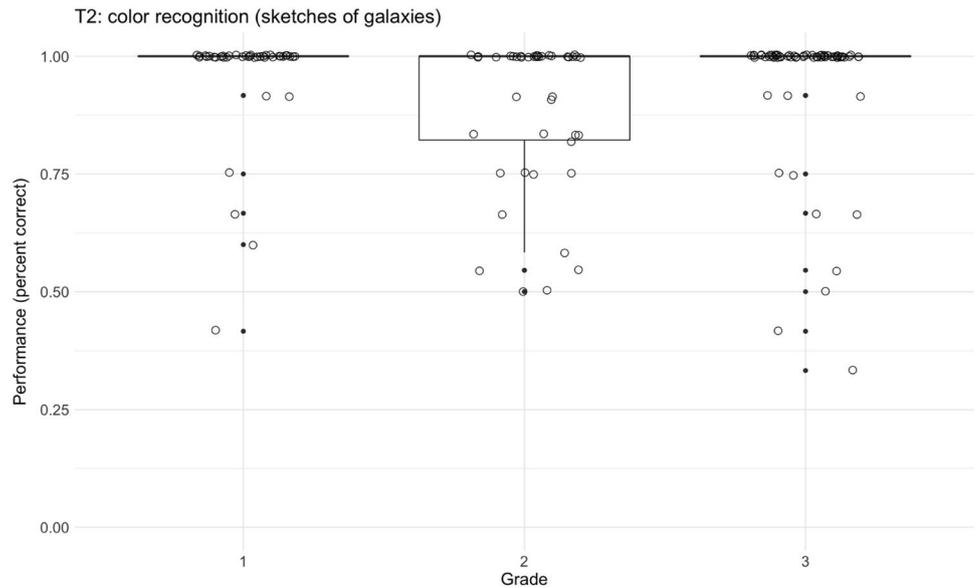

**Fig. 3** Performance of students in the testing phase T2 for colour identification tasks of the sketches of galaxies. Results are represented as a function of the grade of the participants. Adopted symbols are the same as in previous figures

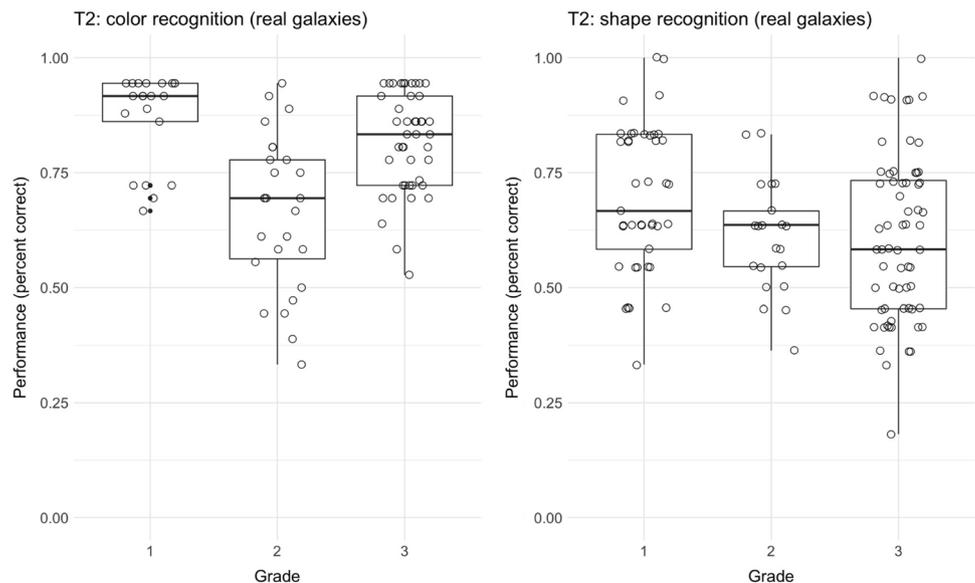

**Fig. 4** Performance of students in the testing phase T2 for colour identification (left) tasks and for the shape identification task (right) of real galaxies. Results are represented as a function of the grade of the participants. Adopted symbols are the same as in previous figures

recognised more than two colours per image: $t(90) = 8.15$, $p < 0.0001$, $d = 1.20$, $BF = 4{,}177{,}246{,}846$. The students' performance was not identical in the three grades, with the second-grade scoring on average lower than first and third-grade students: $F(2, 88) = 18.79$, $p < 0.0001$. The results are reported in Fig. 4 (left).

We also analysed shape recognition of real galaxies. One hundred and twenty-six students completed this task. Students recognised the shape above chance: $t(125) = 9.13$, $p < 0.0001$, $d = 1.15$, $BF = 4.177327e + 12$, and a small performance difference was observed among the grades: $F(2, 123) = 3.88$, $p = 0.0232$ (see Fig. 5, right).

### 4.2 Engagement with scientific disciplines

We analysed the data gathered with the mATSI questionnaire. One hundred and thirty-eight students responded to the questionnaire in Q1, and ninety-one responded in Q2. The drop in the number of respondents was due to the proximity of the end of the school year with many classes that were engaged in the preparation of the exams of the end of the year. Among these responses, we were able to select the responses of sixty-seven students who responded both in Q1 and Q2 and also participated in at least one of the testing sessions.

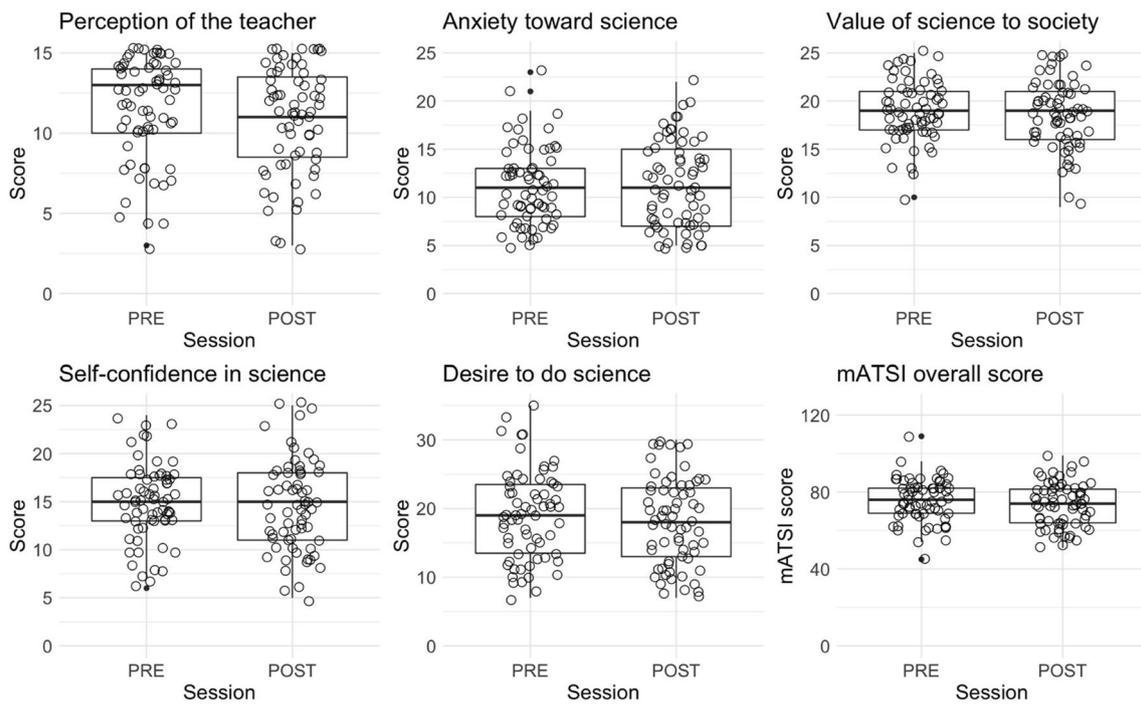

**Fig. 5** Responses of the students to the mATSI questionnaire before (PRE) and after (POST) the testing sessions T1 and T2. The plots show the scores of the individual subscales of the questionnaire and the comprehensive score of the test (bottom right plot)

**Table 3** Results of the paired samples *t*-tests calculated on the mATSI scores collected in Q1 and Q2. Columns report the value of the *t* statistic, the *p*-value, the Cohen's *d*, the Bayes Factor (BF) mean, and the standard deviation (in brackets) of the score recorded in Q1 and Q2

| Score | t(132) | p | Cohen's d | BF | M(SD)-Q1 | M(SD)-Q2 |
| --- | --- | --- | --- | --- | --- | --- |
| Comprehensive mATSI score | 1.15 | 0.25 | 0.19 | 0.70 | 75.6 (11.1) | 73.3 (11.3) |
| Perception of the teacher | 1.27 | 0.20 | 0.22 | 1.02 | 11.4 (3.1) | 10.7 (3.3) |
| Anxiety for science | 0.26 | 0.79 | 0.04 | 0.14 | 11.1 (3.9) | 11.3 (4.5) |
| Value of science | 0.62 | 0.53 | 0.10 | 0.17 | 18.9 (3.1) | 18.5 (3.5) |
| Self-confidence in science | 0.51 | 0.60 | 0.08 | 0.20 | 14.9 (3.9) | 14.5 (4.7) |
| Desire for science | 0.87 | 0.38 | 0.15 | 0.43 | 19.0 (6.4) | 18.0 (6.2) |

The mATSI questionnaire returns one comprehensive score of "science engagement" and five subscores on particular aspects of the relationship between the student and science: the perception of the teacher, the anxiety towards science, the value of science to society, the self-confidence in science, and the desire to do science. The scores (i.e. the comprehensive score and the various subscores) were calculated for each student participating in both Q1 and Q2 and compared by a set of paired sample *t*-tests. The results of these tests are reported in Table 3 and shown in Fig. 5. Overall, the questionnaire revealed no significant change in mATSI scores before and after the testing sessions with Edukoi.

Although we observed no quantitative difference between Q1 and Q2, the verbal feedback that we received from the students was very positive. In class and in the semi-structured interviews conducted after the experiment, students mentioned that recognising colours was easy, whereas recognising shapes was more challenging. Some students also mentioned that this method would be helpful for blind students and that they had the chance to imagine how a blind person feels. They also mentioned that exploring geometric shapes and astronomical images through sound was fun, enjoyable, and interesting. We cannot exclude that these responses reflected social desirability; however, we also received negative feedback from a few students who did not find the experiment particularly interesting. See Appendix 3 for the most significant quotes we recorded.

# 5 Discussion

Our study sought to fill a gap in the literature regarding the efficacy of sonification in communicating astronomical concepts within educational settings. Indeed, while studies have explored the effectiveness of sonification in astronomy research (e.g. Tucker-Brown et al. 2022, Cooke et al. 2019) and outreach (e.g. Harrison et al. 2022, Varano & Zanella 2023), experiments tailored specifically to astronomy in formal education contexts are missing. Here, we discuss our results in light of current literature studies and highlight possible ways forward.

## 5.1 Effectiveness of sonification for colour recognition

Our study showed that colour-recognition tasks had a correct-response rate close to 100% when only one base RGB colour (red, green, blue) was present in the image, and >75% when multiple colours were played at the same time. This suggests that the adopted colour-sound mapping is effective. This appears remarkable given that the training period was very short (about 5 min) and it was only performed during T1. When we repeated the experiment after 2 months of inactivity, the participants still remembered the mapping. This shows that our coding is effective and easy to learn; hence, it is suitable for use in the school context.

Several other projects in recent years have explored different ways of mapping colour into sound and have tested its effectiveness [39]. Often colour is represented by different sound frequencies or different musical instruments. In the latter case, the colour-instrument association is often chosen based on the results of Rossi et al. [40], who, on the statistics of more than 700 people, produced a table summarising the most common associations that individuals produce between colours and musical instruments (e.g. oboe for red, viola for orange, flute for green). Cavaco et al. [41] mapped seven colours into pitch and showed that participants could identify the colour of 51% of the samples after an initial training of about 10 min. Other studies, in contrast, mapped colours to musical instruments (e.g. [42]), yielding success rates ranging between 54 and 81% with training from 45 min up to 2–3 h. Although this mapping seems to be more effective than the colour–pitch code, the training needed to obtain good performances is long. This seems unsuitable for educational environments such as schools, the target of our study. In addition, Capalbo and Glenney [43] showed that only individuals with perfect pitch perform well when this mapping is adopted. However, perfect pitch is extremely rare among the population [44]. For this reason, Capalbo and Glenney [43] suggested using panning and pitch to discern between colours. We found that associating colours with natural sounds rather than musical instruments was intuitive for students and could be easily memorised. From our study, it is unclear whether the panning, which we added as a redundant stimulus, helped identify colours, although we may speculate that it helped when multiple (or composite) colours were present in a picture. We are aware, however, that panning is not always effective, such as when the student is not wearing headphones correctly or s/he is listening to the sonification via loudspeaker and so on. We also noticed that the use of natural sounds instead of musical instruments allowed for prompt recognition of the colours for students who were not used to listening to music and/or had a different cultural background (about 10% of the students had Chinese, Indian, Ukrainian, or African origins) and could therefore be used to different instruments. We did not register significant differences across grades instead, likely because the age range was rather small (11–14 years old).

One limitation, common to studies mapping colour to different musical instruments or other natural sounds (as in our case), is the limited number of colours that can be represented to avoid confusion and difficulties in memorising the associations. The participants' knowledge of additive colour synthesis would help to overcome this issue. However, most of the students, especially in the first grades, may not be familiar with it or do not remember it; thus, we cannot rely on this knowledge at least at these young ages. The use of panning to add information about additional colours, instead of being redundant as it was in our study (green at the centre and with the sound of birds; red to the right and with the sound of fire; blue to the left and with the sound of bubbles), might help to represent additional secondary colours. For example, following Capalbo and Glenney [43], both blue and yellow could be mapped to the left, but with different sounds. Similarly red and green could be mapped to the right, while black and white to the centre. However, it remains critical how one would distinguish between colours with similar RGB or CMYK values/proportions, such as white or yellow in RGB coding; loudness is not an efficient parameter in sonification [45], thus, and it would be necessary to manipulate other dimensions of sounds. A possible solution could be mapping more colours to different sounds. In this scenario, it would be necessary to find sounds that are easily associated with each colour but dissimilar in spectral characteristics (e.g. with non-overlapping spectra) so that a listener can easily distinguish them.

## 5.2 Effectiveness of sonification for shape recognition

Although students recognised colours easily, their capability to identify geometric shapes was less robust, albeit above the chance level both for geometric/stylised shapes and for real galaxy shapes. The students put in place different strategies to recognise shapes: some were exploring the space in an ordered way, for example, from left to right and/or from top to bottom, and others were going in circles (especially when they were asked to distinguish between elliptical and spiral shapes), whereas others were exploring the space randomly. We could not assess if there was a more effective strategy, but this could also depend on the specific learning preference of each student. Definitely, one limit of our approach is the direct translation of individual pixels into sound that is not always efficient in allowing to capture the gestalt of the object the user is exploring [45, 46].

In the literature, we find several studies on the topic of shape sonification with various sound mappings [42, 47–49]. These sonification techniques are based on a scanning approach; the user cannot choose how to explore the images. The studies that involve the possibility of interacting and customising the sonification are rarer. Some experiments show how shape recognition through devices with a tactile component, such as tablets, has recognition rates higher than 60% with different sonification mappings [50]. However, the accuracy of discrimination depends on the complexity of the shape and how similar it is to other presented shapes [51]. Despite the fact that the galaxy images that we presented to the students could appear quite similar (e.g. the bright central area of spirals could resemble the diffuse halo of ellipticals, see Fig. 1 and Appendix 2), we obtained average performances as high as 60%, indicating that leaving the user free to explore the image, without forcing a single approach (e.g. scanning approach) is effective. Performances might become even higher by adding a support or tactile frame to guide the hand of the participants, such as a transparent (e.g. plexiglass) surface that allows the webcam to recognise the hand of the participant and—at the same time—provide the user with a visual and haptic reference about the hand placement in the virtual, two-dimensional space. This would also solve an issue that occurred several times, especially during T1: some students tended to move their hand towards the webcam instead of in front of it. This was causing problems in the hand recognition by the webcam, and it was making the image exploration, and in turn the shape recognition, more difficult. Implementing a tablet version of the Edukoi might also help to overcome these challenges, as there is evidence that sonification implemented in tablets allows high recognition rates of stylised shapes even with minimal training [50].

## 5.3 Effectiveness of sonification for engaging students with STEM

In our study, we used the mATSI questionnaire to understand if our interventions could positively impact students' attitudes towards STEM subjects. The results showed no difference in the mATSI scores before and after the intervention. This is not surprising, given the relatively small size of the intervention presented here (i.e. only two sessions of 1 h per student over an entire school year). The oral feedback of the students recorded at the end of the sessions might be more informative. Students enjoyed Edukoi and gave positive feedback about its use in the class and in the brief personal interviews that were conducted later in the day of T2. They stressed several times that the tool was fun and that it was a useful and unexpected surprise within the science class. Several students highlighted that they would like to use sonification and Edukoi more often at school. There were also several comments about how using a sonification tool helped them to understand how their BVI peers feel in everyday life.

The literature about the effectiveness of sonification in the study of STEM subjects [52] and, especially in middle schools [21, 53], is slowly growing. Christidou [54] argued that there is a loss of interest in science as students advance from primary to secondary education. During middle and high school, many students cease seeing scientific disciplines as a viable option for their future. The decline in students' interest, motivation, or attitude towards certain subjects can be attributed to various factors [16]. One of the significant factors is self-efficacy. Accumulated positive experiences in the domains of science and technology, such as extracurricular activities or even initiatives within the school, can potentially deter these declines [16]. The contextualisation of content is imperative to captivate the interest of students. This pedagogical approach, which involves integrating educational content with real-world applications and phenomena, has been recurrently advocated for in existing literature [54, 55]. For instance, our study linked science education with an actual astronomical application, such as the investigation of galaxy colours and shapes—that, in turn, helps astronomers to understand galaxy formation and evolution. Highlighting such links with real-life applications could benefit the students' overall perception of science and technology.

In addition, creating a collaborative endeavour that engages and integrates novel ideas is important to maintain the interest of students in scientific disciplines [55]. Our study, which involved an interactive tool (Edukoi) and a new technique (sonification) to learn about science (together with the meeting with a practising astronomer), goes exactly in this direction. Students' enthusiasm was evident in the oral feedback we received in the classroom when asked to freely comment about their experience with Edukoi, but also in the semi-structured interviews. Some quotes are "It was a lot of fun."; "It was amazing and difficult at the same time."; "I enjoyed it very much. It was fun and enjoyable, and at the same time complicated and interesting. I would like to do it again."; "It was unexpected to observe with a computer through sound, but it was fun, and I hope to do it again."; "I learned patience and persistence, things I never had before. But there [on Edukoi] going randomly doesn't help, so you have to follow certain paths."; "I've learned that I'm starting to like Space." (more quotes are reported in Appendix 3). They highlight how positive this experience was for the students and how such interactive and multi-sensory lessons seem to be an effective way of teaching science. Of course, we cannot exclude that these comments reveal social desirability. In addition, the fact that students kept their attention for the whole administration of behavioural tests, despite some tasks (e.g. shape recognition) might be challenging and frustrating at times, shows how sonification and interactive teaching are not just a playful activity and could instead allow teachers to achieve also demanding tasks.

### 5.4 Sonification for the accessibility of science by blind and visually impaired people

Currently, there is no golden standard to convey colour information to BVI people, although several alternatives are being investigated [56]. Braille-related systems have the limit of relying on memory alone to convey the colour [56]. When these tools allow the free exploration of the images, they present problems that are analogous to Edukoi, such as the risk of "getting lost" in the image. By the same token, BVI people appreciate the possibility of exploring images, and sonification might be a way to implement this exploration easily, without the need for ad hoc coding such as in Braille [56]. During the interviews, several students spontaneously highlighted how the use of sound could help blind students to attend science classes. This shows how the use of sonification in science classes is not only helpful in conveying information and contents but also in widening the perspectives and imagination of sighted students in the context of accessibility.

## 6 Summary and conclusions

We conducted a study to assess sonification's effectiveness in conveying information (i.e. colour, shape) of geometric shapes and astronomical images, as well as the impact that sonification might have in engaging students in scientific disciplines. We adopted the interactive sonification tool Edukoi and we mapped RGB colours to natural sounds (red = fire, green = birds and forest, blue = water bubbles). This mapping was aimed at easing memorisation and recollection even after a time of inactivity, and it was also meant to overcome cultural and background barriers or preconceptions that might instead be linked to the use of timbres of musical instruments or pitch.

We ran the experiment in a middle school with about 150 students aged 11–14, including one blind student, three students with significant sensitive or cognitive disabilities, and another ten students with mild cognitive disabilities. About 10% of the students were not Italian, although most had good Italian proficiency.

We used two evaluation tools: (1) the mATSI questionnaire [17, 31], translated to Italian, to evaluate the engagement of the students, and (2) a series of colour and shape recognition tasks to be performed by using the Edukoi software. Tests were performed in the classroom, and each student had a laptop with the screen covered by a black and opaque sheet of paper preventing them from seeing the images, to evaluate the efficacy of sonification only. We administered the mATSI questionnaire and ran semi-structured interviews before (Q1) and after (Q2) the colour and shape recognition tasks. The colour and shape recognition tasks were run twice (T1 and T2) with a time interval of about 2 months.

We found that:

- All students had a good success rate in colour recognition with an average performance > 85% and a median performance of 100% both in T1 and T2 when sketchy images were involved. The performance in colour recognition did not significantly depend on the students' school grade, and it was very good also when real astronomical images with complex colour patterns were explored (median success rate > 80%).
- While in T1 we explained the colour-to-sound mapping adopted by Edukoi, in T2, we let the students recall it. The excellent performance, almost at ceiling, in the colour recognition of sketchy galaxies (that presented only presented one colour at time) in T2 (93%) shows that the majority of students effectively recalled the colour mapping, irrespective of their school grade (i.e. age), and likely irrespective of their cultural background.
- The success rate in the shape recognition tasks was above chance level, with average performances of 45% in T1 for geometric shapes and 64% in T2 for real galaxies. Performance did not significantly depend on the students' school grade.
- The score of the mATSI test before and after our intervention in the school was comparable, indicating no substantial changes in the engagement of students after the introduction of sonification and the use of Edukoi. The oral feedback that the students gave us after the use of Edukoi however was positive: they enjoyed using sound to explore geometric and scientific images, they found it useful although difficult at times (especially the shape recognition), they felt surprised by the tool, and they expressed the will of using it more often.

Overall, our study shows the potential of sonification in conveying specific information when teaching science classes. Sonification has proven quite successful not only when adopted to explore rather simple geometric shapes (square, triangle, circle) and monochromatic primary colours (red, green, blue), but also with realistic and more complex images (e.g. actual astronomical observations). Although many improvements in the sonification software are possible (e.g. the use of a support or tablet that could be touched while exploring images to help guide hands in shape recognition), Edukoi has proved to be a fun tool, versatile, and effective in the limits of the tasks proposed.

## Appendix 1. The mATSI questionnaire

We administered the mATSI questionnaire twice, at the beginning (Q1) and at the end (Q2) of our study. Being all the students Italian speakers, we translated the original English questionnaire to Italian. In the following, we report the original questions in English and our Italian translation. Answers were given on a Lickert scale from 1 (not at all) to 5 (very much).

- Science professors make science interesting
    *I professori di Scienze rendono la scienza interessante*
- Science teachers present material in an interesting way
    *I professori di Scienze spiegano gli argomenti in modo interessante*
- Science teachers are willing to give us individual help
    *Se uno di noi non capisce qualcosa, i professori di Scienze glielo spiegano volentieri*
- It makes me nervous to even think about doing science
    *Anche solo pensare alle materie scientifiche mi rende nervoso/a*
- I feel tense/nervous when someone talks me about science
    *Quando qualcuno mi parla di materie scientifiche mi sento a disagio*
- It scares me to have to take a science class
    *Fare Scienze a scuola mi spaventa*
- When I hear the word "science", I have a feeling of dislike
    *Quando sento la parola "scienze", provo una sensazione di disgusto*
- I have a good feeling toward Science
    *Le materie scientifiche mi piacciono*
- Science is useful for solving the problems of everyday life
    *La scienza è utile per risolvere i problemi della vita di tutti i giorni*
- Science is helpful in understanding today's world
    *La scienza serve a capire come funziona il mondo*
- Most people should study some science
    *Tutti dovrebbero studiare un po' le materie scientifiche*
- Science is of great importance to a country's development
    *La scienza è molto importante per lo sviluppo di un Paese*
- It is important to know science in order to get a good job
    *Conoscere le materie scientifiche è importante per trovare un buon lavoro*
- No matter how hard I try, I cannot understand science
    *Per quanto io ci provi, non riesco a capire le materie scientifiche*
- Science is easy for me
    *Le materie scientifiche sono facili per me*
- I usually understand what we are talking about in science
    *Di solito capisco di cosa si parla quando si parla di scienza*
- I do not do very well in science
    *Non sono bravo/a in Scienze*
- I often think, "I cannot do this", when a science assignment seems hard
    *Spesso penso: "Non ci riesco", quando un compito di Scienze sembra difficile*
- Science is something which I enjoy very much
    *Mi piacciono molto le materie scientifiche*
- I like the challenge of science assignments
    *Mi diverto a svolgere i compiti di Scienze*
- I have a real desire to learn science
    *Mi piace davvero molto studiare le materie scientifiche*
- Science is one of my favorite subjects
    *Scienze è una delle mie materie preferite*
- I would like to do some reading in science which has not been assigned to me
    *Mi piacerebbe leggere più materiale di quello che mi è stato assegnato in Scienze*
- Sometimes I read ahead in our science book
    *A volte leggo più avanti nel libro di Scienze di dove siamo arrivati a lezione*
- It is important to me to understand the work I do in the science class
    *Per me è molto importante capire quello che facciamo nelle ore di Scienze*

## Appendix 2. Sketches and images of galaxies used in T2

In T2, the students were presented with twelve sketches of galaxies (Fig. 6) and twelve images of actual spiral and elliptical galaxies (Fig. 7).

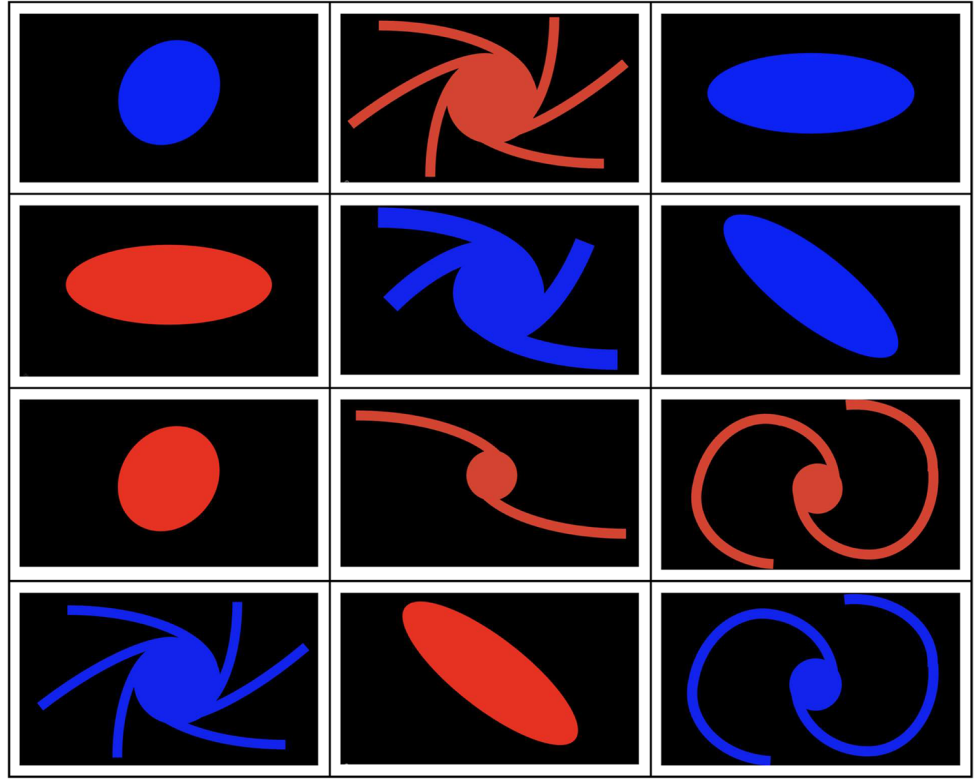

**Fig. 6** Sketches presented during the first task of the T2 session

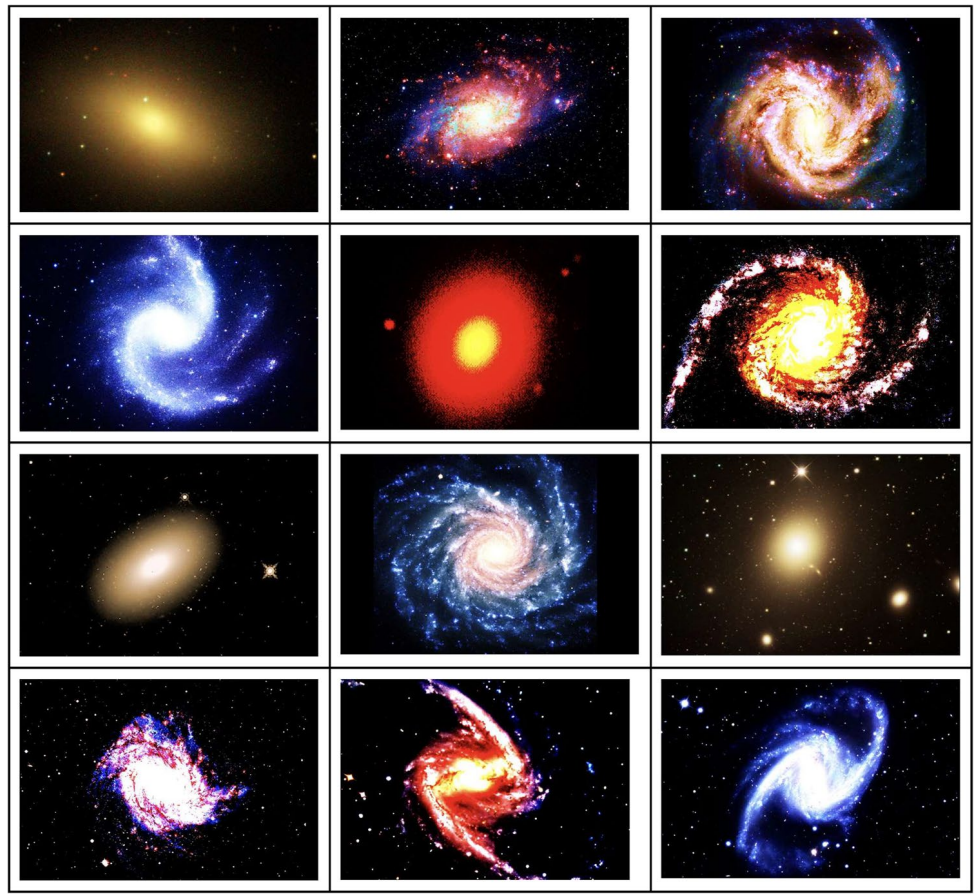

**Fig. 7** Images of galaxies presented during the second and third task of the T2 session. From top to bottom and from left to right, we adopted images of the following sources: Messier 33 (observed with the VLT Survey Telescope), NGC 1232 (observed with FORS on the VLT), Markarian 1216 (observed with the Hubble Space Telescope), M32 (observed with the 1.1 Meter Hall Telescope), M87 (observed with the Hubble Space Telescope), M61 (observed with FORS on the VLT), M59 (observed with the Hubble Space Telescope), M83 (observed with the ESO 2.2 m telescope), NGC 1365 (observed with the Hubble Space Telescope), NGC 5247 (observed with HAWK-I on the VLT), NGC 1566 (observed with the Hubble Space Telescope), NGC 1365 (observed with the Hubble Space Telescope). We adjusted the contrast and saturation of the images

# Appendix 3. Quotes from the students about the use of sonification in the class

The students were asked to comment freely about how they felt about using sound and Edukoi to explore geometric shapes and astronomical images during science lessons. Here, we report some of the most significant quotes (the original Italian quote is in bracket.

- It was a lot of fun. [*È stato molto divertente.*]
- It was amazing and difficult at the same time. [*È stato molto bello e difficile allo stesso tempo.*]
- It was difficult, but now I understand how blind people can study stars. It was also fun. [*È stato difficile, ma adesso so come fanno i non vedenti a studiare le stelle. È stato anche divertente.*]
- It was beautiful. [*È stato bello.*]
- It was beautiful, but also difficult. [*È stato bello, ma anche difficile.*]
- I enjoyed it very much. It was fun and enjoyable, and at the same time complicated and interesting. I would like to do it again. [*Mi è piaciuto molto. È stato divertente e piacevole, e allo stesso tempo complicato e interessante. Lo rifarei.*]
- This project was very interesting and fun. Space and science are very educational and I like them. This is an experience that I would like to repeat multiple times. To be honest, it was neither easy nor difficult. [*Questo è un progetto molto interessante e divertente. Lo spazio e la scienza sono molto educativi e mi piacciono. Questa è un'esperienza che vorrei rifare molte volte. Se devo essere sincero non era nè facile nè difficile.*]
- It was interesting. I was surprised to learn about colours and shapes through sound. It was very interesting and it needs a lot of concentration. I would like to do it again. [*È stato interessante. Mi ha sorpreso capire i colouri e le forme in base ai suoni. È stato molto interessante e c'è bisogno di molta concentrazione. Lo rifarei di nuovo.*]
- Very interesting, especially the sounds associated with colours. It was a beautiful experience also to imagine a different reality. [*Molto interessante, soprattutto i suoni associati ai colouri. È stata una bella esperienza, anche immedesimarsi in una realtà diversa.*]
- I liked it, it was beautiful and interesting and the sounds that the computer produced were very satisfactory. I would like to do it again. [*Mi è piaciuto, è stato bello e interessante e i suoni che produceva il computer erano molto soddisfacenti. Mi piacerebbe rifarlo.*]
- I think it was useful, but I would like it to be easier. [*Secondo me era utile, però lo rifarei un po' più semplice.*]
- It was unexpected to observe with a computer through sound, but it was fun and I hope to do it again. [*Non me l'aspettavo di osservare dal computer diversi suoni, ma mi sono divertito e spero di rifarlo.*]
- It was beautiful, but weird. It was very confusing at times. But it was also beautiful. [*È stato bello, ma strano. È stato molto confusionario a volte. Ma è stato anche bello.*]
- I had never thought that [blind people] could have difficulties with electronic things, but now I know. [*Non avevo mai pensato che [le persone non vedenti] potessero avere difficoltà nelle cose elettroniche, mentre adesso lo so.*]
- It was fun, but also serious and educational. It's great that with technology we were able to achieve a result for people who cannot see. [*È stato divertente, ma anche serio ed istruttivo. Bello che con la tecnologia siamo riusciti ad avere un risultato per le persone che non ci vedono.*]
- I learned patience and persistence, things I never had before. But there [on Edukoi] going randomly doesn't help, so you have to follow certain paths. [*Ho imparato calma e costanza, cose che non ho mai avuto. Ma lì [su Edukoi] andare a casaccio non aiuta, e quindi bisogna seguire dei percorsi.*]
- Now I know that even the blind can become scientists; I had my doubts before. [*Adesso so che anche i ciechi possono diventare scienziati, prima avevo il dubbio*]
- I've learned that I'm starting to like Space. [*Ho imparato che lo Spazio inizia a piacermi*]

**Acknowledgements** We would like to thank the reviewers for their time, effort, and fruitful criticism. We thank the students, teachers, and staff from the Scuola Media Don Milani who supported our first testing phase. In particular, we thank the teacher Roberto Veltri and the psychologist Arianna Daldosso for their practical support during the project.

**Author contribution** Conceptualization: Zanella A, Guiotto Nai Fovino L, Grassi M; methodology: Grassi M, Guiotto Nai Fovino L, Zanella A; formal analysis and investigation: Grassi M, Guiotto Nai Fovino L, Zanella A; writing—original draft preparation: Guiotto Nai Fovino L, Zanella A, Grassi M; writing—review and editing: Zanella A, Guiotto Nai Fovino L, Grassi M; software development: Di Mascolo L, Ginolfi M; data acquisition: Guiotto Nai Fovino L, Carpita N, Trovato Manuncola F; resources: University of Padua, Istituto Nazionale di Astrofisica; supervision: Zanella A, Grassi M.

**Funding** Open access funding provided by Università degli Studi di Padova within the CRUI-CARE Agreement. This work has been supported by the French government, through the UCA[J.E.D.I.] Investments in the Future project managed by the National Research Agency (ANR) with the reference number ANR-15-IDEX-01.

**Data availability** Data and analysis scripts are available in the Open Science Framework.

## Declarations

**Conflict of interest** The authors declare no competing interests.